\documentclass[mathleft
]{an}
\usepackage{graphicx}
\usepackage{times}
\usepackage{amssymb}
\def\lapp{\ifmmode\stackrel{<}{_{\sim}}\else$\stackrel{<}{_{\sim}}$\fi}
\def\gapp{\ifmmode\stackrel{>}{_{\sim}}\else$\stackrel{>}{_{\sim}}$\fi}
\usepackage{color}

\begin{document}

\Pagespan{1}{}
\Yearpublication{2014}%
\Yearsubmission{2013}%
\Month{8}%
\Volume{999}%
\Issue{88}%

\title{NuSTAR results and future plans for magnetar and rotation-powered pulsar observations}

\author{Hongjun An\inst{1},
Victoria M. Kaspi\inst{1}\fnmsep\thanks{Lorne Trottier Chair; Canada Research Chair},
Robert Archibald\inst{1},
Matteo Bachetti\inst{2,3}, Varun~Bhalerao\inst{4,5}, Eric~C.~Bellm\inst{4}, Andrei~M.~Beloborodov\inst{6}, Steven E. Boggs\inst{7},
Deepto Chakrabarty\inst{8}, Finn~E.~Christensen\inst{9}, William~W.~Craig\inst{7,10},
Fran{\c c}ois Dufour\inst{1}, Karl Forster\inst{4}, Eric V. Gotthelf\inst{6}, Brian~W. Grefenstette\inst{4},
Charles J. Hailey\inst{6}, Fiona A. Harrison\inst{4}, Romain Hasco{\"e}t\inst{6},
Takao Kitaguchi\inst{11}, Chryssa Kouveliotou\inst{12}, Kristin K. Madsen\inst{4},
Kaya Mori\inst{6}, Michael J. Pivovaroff\inst{10}, Vikram~R.~Rana\inst{4},
Daniel~Stern\inst{13}, Shriharsh Tendulkar\inst{4}, John A. Tomsick\inst{7}, Julia K. Vogel\inst{10},
William W. Zhang\inst{14}, and the NuSTAR Team\\}
\titlerunning{{\em NuSTAR} results and future plans for magnetar and rotation-powered pulsar observations}
\authorrunning{An et~al.}
\institute{
{\small $^1$Department of Physics, McGill University, Montreal, Quebec, H3A 2T8, Canada}\\
{\small $^2$Universit{\'e} de Toulouse, UPS-OMP, IRAP, Toulouse, France}\\
{\small $^3$CNRS, Institut de Recherche en Astrophysique et Plan{\'e}tologie, 9 Av. colonel Roche, BP 44346, F-31028 Toulouse Cedex 4, France}\\
{\small $^4$Cahill Center for Astronomy and Astrophysics, California Institute of Technology, Pasadena, CA 91125, USA}\\
{\small $^5$Inter-University Center for Astronomy and Astrophysics, Post Bag 4, Ganeshkhind, Pune 411007, India}\\
{\small $^6$Columbia Astrophysics Laboratory, Columbia University, New York NY 10027, USA}\\
{\small $^7$Space Sciences Laboratory, University of California, Berkeley, CA 94720, USA}\\
{\small $^8$Kavli Institute for Astrophysics and Space Research, Massachusetts Institute of Techolology, Cambridge, MA 02139, USA}\\
{\small $^9$DTU Space, National Space Institute, Technical University of Denmark, Elektrovej 327, DK-2800 Lyngby, Denmark}\\
{\small $^{10}$Lawrence Livermore National Laboratory, Livermore, CA 94550, USA}\\
{\small $^{11}$RIKEN, 2-1 Hirosawa, Wako, Saitama, 351-0198, Japan}\\
{\small $^{12}$Space Science Office, ZP12, NASA Marshall Space Flight Center, Huntsville, AL 35812, USA}\\
{\small $^{13}$Jet Propulsion Laboratory, California Institute of Technology, Pasadena, CA 91109, USA}\\
{\small $^{14}$Goddard Space Flight Center, Greenbelt, MD 20771, USA}\\
}

\received{24 August 2013}
\accepted{XXX}
\publonline{YYY}


\abstract{The {\em Nuclear Spectroscopic Telescope Array (NuSTAR)} is the first focusing hard X-ray mission
in orbit and operates in the 3--79 keV range. {\em NuSTAR}'s sensitivity is roughly two orders of magnitude
better than previous missions in this energy band thanks to its superb angular resolution.
Since its launch in 2012 June, {\em NuSTAR} has performed excellently and observed many interesting sources
including four magnetars, two rotation-powered pulsars and the cataclysmic variable AE~Aquarii.
{\em NuSTAR} also discovered 3.76-s pulsations from the transient source SGR~J1745$-$29 recently found
by {\em Swift} very close to the Galactic Center, clearly identifying the source as a transient magnetar.
For magnetar 1E~1841$-$045, we show that the spectrum is well fit by an absorbed blackbody
plus broken power-law model with a hard power-law photon index of $\sim$1.3. This is consistent with
previous results by {\em INTEGRAL} and {\em RXTE}. We also find an interesting double-peaked pulse profile
in the 25--35 keV band. For AE~Aquarii, we show that the spectrum can be described by a multi-temperature
thermal model or a thermal plus non-thermal model; a multi-temperature thermal model without a non-thermal
component cannot be ruled out. Furthermore, we do not see a spiky pulse profile in the hard X-ray band, as previously
reported based on {\em Suzaku} observations. For other magnetars and rotation-powered pulsars observed with
{\em NuSTAR}, data analysis results will be soon available.}

\maketitle

\section{Introduction}
A neutron star is the left-over dense and hot stellar core of a massive star which survived after the
energetic explosion of a star. Neutron stars typically have strong surface magnetic
fields ($10^9$--$10^{15}\ \rm G$)
inferred from their spin properties. They are typically identified with pulsations in the radio, X-ray,
or gamma-ray band which
are caused by the rapid rotation of the star. There are many subclasses in the neutron star population such as
rotation-powered pulsars (RPPs), rotating radio transients (RRATs), high-$B$ RPPs,  X-ray dim
isolated neutron stars (XDINS), central compact objects (CCOs), and magnetars (see \cite{k10} for
a review).

Radiation properties of neutron stars are diverse. For example, RPPs are mostly observed in the radio
band, but sometimes in the X-ray to gamma-ray band as well. Magnetars typically emit thermal
photons (modified in the magnetosphere) in the soft X-ray band, but several show a
striking turnover and rising power-law emission (in the $\nu F_{\nu}$ representation) in the hard X-ray band.
X-ray properties of X-ray emitting neutron stars have been studied
with the soft X-ray observatories which operate in the $\sim$0.3--10 keV band
(e.g., {\em XMM-Newton} and {\em Chandra}).
However, the hard X-ray emission ($\gapp$10 keV) has not been very well studied because until recently
hard X-ray observatories could only detect a small handful of very bright sources.

While soft-band spectra have been well measured, they are strongly affected by the
hydrogen absorption, and the current results based on the soft-band observations
may be inaccurate. In particular, constraining a non-thermal spectral component is difficult
because it can extend down to low energy
and covaries with the hydrogen column density ($N_{\rm H}$).
Therefore, hard-band observations ($\gapp$10 keV) which
are almost free from the absorption can help us accurately determine the non-thermal spectrum.

\label{sec:targets}
\begin{figure}
\includegraphics[width=80mm,height=60mm]{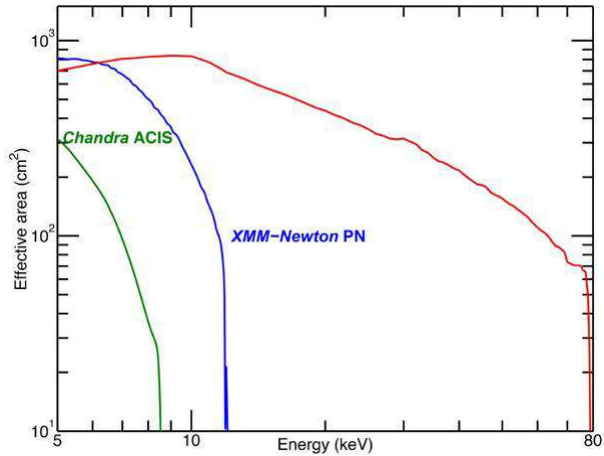}
\caption{Effective area of {\em NuSTAR} (red) together with that of {\em XMM-Newton}/PN (blue) and {\em Chandra}/ACIS (green).}
\label{fig:effarea}
\end{figure}
\medskip

\section{The Nuclear Spectroscopic Telescope Array (NuSTAR)}
{\em NuSTAR} is the first focusing hard X-ray telescope in orbit, operating in the 3--79 keV
band. It is composed of two hard X-ray optics and two focal plane modules (\cite{hcc+13}).
The optics have an effective area of 900 cm$^2$ (Fig.~\ref{fig:effarea}), and
the angular resolution in half power diameter (HPD) is 58$''$ at $\sim$10~keV,
which makes {\em NuSTAR} the most sensitive observatory in the $\sim$8--79 keV band.
The timing resolution of the detectors is 2 $\mu \rm s$, but the accuracy is a little worse
due to the clock drift (3 ms absolute, 100 $\mu \rm s$ relative), and the spectral resolution is
0.4~keV FWHM at 10 keV. {\em NuSTAR} can access 80\% of sky at any time, and its Target-of-Opportunity (ToO)
response is 6--8 hours typically. See Table~\ref{ta:performance} for more details.

{\em NuSTAR} was launched into a 600 km near-circular orbit at 6$^\circ$ inclination on 2012 June 13,
and started its science observations two months after the launch.
Extensive in-orbit calibration was performed, and
its on-orbit performance is similar to what was estimated on the ground.
The first public release of {\em NuSTAR} data, together with calibration files and pipeline tools,
is scheduled for late Summer 2013.

The primary science goals of {\em NuSTAR} are to locate massive black holes in the Universe,
to understand supernovae explosion mechanisms, to study the most powerful cosmic accelerators,
and to survey the center of our Galaxy. In addition
to these, {\em NuSTAR} has many other interesting science goals and constructed science working groups according
to these goals.\footnote{see http://www.nustar.caltech.edu/for-astronomers/science-working-groups}

\begin{table}
\centering
\caption{Performance parameters of {\em NuSTAR}.}
\label{ta:performance}
\begin{tabular}{ccc}\hline
Property & Performance & Comments  \\
\hline
FoV    & $10'$  & 50\% response at 10 keV \\
Angular resolution & 58$''$  & HPD \\
                   & 18$''$  & FWHM \\
Timing accuracy & 100 $\mu$s  & Relative \\
        & 3 ms  & Absolute \\
Spectral resolution  & 0.4 keV & FWHM @ 10 keV \\
ToO response  & $\lapp$24 hours &  \\
\hline
\end{tabular}
\end{table}

In this paper, we focus on magnetars and RPPs which were or will be studied by the
magnetars and RPP working group of the {\em NuSTAR} team.
The target list includes five magnetars, two RPPs, and a unique white dwarf binary.
In Section~\ref{sec:targets}, we present an overview of the target list and the observations.
We then show the data analysis results for the objects already observed in Section~\ref{sec:result}.
Finally, we summarize and discuss future plan in Section~\ref{sec:concl}.

\section{Magnetars and Neutron Stars with {\em NuSTAR}}
\label{sec:targets}
{\em NuSTAR} is planning on observing magnetars and rotation-powered pulsars for a total of $\sim$ 1.2 Ms with
an emphasis on magnetars. Table~\ref{ta:targetlist} shows the priority A targets.

The main goal of magnetar observations is to understand the hard X-ray emission. Some magnetars are clearly
detected in the hard X-ray band ($\gapp$ 10 keV) while others are not (\cite{khh+06}). It is not clear
if all magnetars emit hard X-rays but some of them may have not been detected due to the sensitivity limitations
of previous hard X-ray observatories.
Alternatively, it is possible that the hard X-ray detected sources are somehow unusual.
Furthermore, the hard X-ray emission mechanism is not yet clearly understood. There are several different
models that discuss the origin of hard emission from magnetars (\cite{hh05}; \cite{bh07}; \cite{bt07}).
Recently, Beloborodov~(2013)
proposed a detailed model which attributes the hard X-rays to the relativistic outflow near the neutron star.

{\em NuSTAR} is well suited for studying magnetars in the hard X-ray band.
Two magnetars 1E~2259$+$586 and 1E~1048$-$5937 in Table~\ref{ta:targetlist} have not yet been clearly detected in the hard band above
$\sim$15 keV. {\em NuSTAR} was able to detect them for the first time
and characterize their spectral properties. The list also includes
hard X-ray bright magnetars 1E~1841$-$045 and 4U~0142$+$61 which have been relatively well studied with {\em RXTE} and
{\em INTEGRAL} (\cite{khh+06}). With the {\em NuSTAR} observations, we are aiming to study the hard-band
spectral properties using the physical model of Beloborodov~(2013).

We will also study transient cooling of magnetars using a ToO program.
Magnetars sometimes show dramatic increases of
persistent emission. After such events, the flux relaxes to its quiescent level on time scales of days
to months. It has been suggested that such relaxation is caused by the passive cooling of
the hot crust (\cite{let02}),
or by untwisting of the external fields (\cite{bt07}; \cite{b09}).
Such an event occurred near the Galactic Center in 2013 April (\cite{drm+13}).
{\em NuSTAR} discovered 3.76-s pulsations as well as a spinning-down of the source
and hence identified the source as a magnetar (\cite{mgz+13}).
{\em NuSTAR} is monitoring the new magnetar SGR~J1745$-$29 in the Galactic Center to determine its spectral
and temporal properties, and to study their evolution (\cite{k14}).

\begin{table}
\caption{{\em NuSTAR} target list for the magnetar and RPP working group.}
\label{ta:targetlist}
\begin{tabular}{cccc}\hline
Source & Exposure & Type & Obs. date \\
       & (ks) &  &  \\
\hline
AE Aquarii & 126 & White Dwarf & 2012 Sep. \\
Geminga    & 50  & RPP & 2012 Sep. \\
1E~1841$-045$  & 45  & Magnetar & 2012 Nov. \\
SGR~J1745$-$29 & 150 (ToO)  & Magnetar & 2013 Apr. \\
1E~2259$+$586  & 170  & Magnetar & 2013 Apr. \\
PSR~J1023$+$0038  & 100  & RPP & 2013 June \\
1E~1048$-$5937  & 400  & Magnetar & 2013 July \\
4U~0142$+$61  & 100  & Magnetar & T.B.D \\
\hline
\end{tabular}
\end{table}

We are also studying two RPPs, Geminga and PSR~J1023$+$0038. Geminga is a gamma-ray bright pulsar
whose spectrum has been well measured from the IR to TeV band.
The soft X-ray ($\lapp$10 keV) spectrum was well measured with
{\em XMM-Newton}. However, the soft X-ray spectrum does not seem to connect to the optical or to the
GeV spectrum, implying there must be a spectral break (\cite{kpz+05}).
We are searching for a spectral break in the 3--79 keV band using {\em NuSTAR}.
PSR~J1023$+$0038 is a transient object between a Low Mass X-ray Binary (LMXB) and a RPP (\cite{asr+09}).
This object will give us a unique opportunity to understand the hard X-ray emission mechanism
from LMXBs and its relationship to RPP emission (\cite{bdd+03}).

Finally, we are also planning to observe a unique white dwarf system. AE~Aquarii is an
Intermediate Polar (IP) which was shown to emit non-thermal X-rays and a peculiar pulsation in {\em Suzaku}
observations (\cite{thi+08}). Based on such observed features, it was suggested that AE~Aquarii may
accelerate particles in its magnetosphere as RPPs do. However, the hard-band detection made with {\em Suzaku}
was marginal, and a much better detection is required to clearly tell if the source really acts like a RPP.

\section{Data Analysis Results and Current Status}
\label{sec:result}
As of 2013 August, {\em NuSTAR} observed all the priority A targets of the magnetar and RPP working
group except for 4U~0142$+$61 (Table~\ref{ta:targetlist}) and intensive data analyses are on-going.
In this section, we show the current analysis results and status for the observed sources.

\subsection{SGR~J1745$-$29}
SGR~J1745$-$29 is a recently discovered magnetar in the direction towards the Galactic Center.
Following a magnetar-like burst detected with {\em Swift} (\cite{drm+13}), {\em NuSTAR} observed the source.
Mori et~al. (2013)
discovered the 3.76-s pulsation, and measured the spectral and temporal properties of the source
to identify the source. The spin-down rate was measured to be $(6.5\pm 1.4)\times 10^{-12}\ \rm s\ s^{-1}$,
implying a spin-inferred magnetic field $B=1.6\times 10^{14}\ \rm G$.
The {\em NuSTAR}- and {\em Swift}-measured spectrum was well fit to an absorbed blackbody plus power-law model
having $kT\sim 1\ \rm keV$, $\Gamma \sim 1.5$, and 2--79 keV luminosity of $3.5\times 10^{35}\ \rm erg\ s^{-1}$
for an assumed distance of 8 kpc. With the measured spectral and temporal properties, the source was
clearly identified as a transient magnetar.

Since the discovery, we have been monitoring the source using a ToO program
to measure its spectral/temporal properties and their evolution, and to study the transient cooling (\cite{k14}).
The source has been cooling very slowly, and
as of today, the flux is still much higher than the quiescent value inferred from a non-detection
in the {\em Chandra} survey (\cite{mbb+09}).

\subsection{1E~1841$-$045}
1E~1841$-$045 was one of the brightest magnetars in the hard X-ray band reported by Kuiper et~al. (2006).
The hard X-ray bright magnetars typically show a turnover at $\sim$10 keV, which can be very well studied with
{\em NuSTAR} because it is most sensitive at that energy. By accurately measuring the spectrum near the
turnover, we can test the hard X-ray emission model of Beloborodov~(2013).

The previous measurements of the
hard-band photon index made with {\em Suzaku} (\cite{mks+10}) and {\em RXTE}/{\em INTEGRAL} (\cite{khh+06})
agree only marginally in the hard band.
Using our {\em NuSTAR} data, we find that the spectrum is consistent
with the results of Kuiper et~al.~(2006),
but not with those of Morii et~al. (2010).
This may be due to the imperfect Kes~73 background subtraction in the {\em Suzaku} data.
We also found that the pulse profile in the 24--35 keV band shows an interesting double-peaked shape unlike
in the other energy bands. Such deviation in a narrow energy range may suggest
a possible absorption or emission feature in the spectrum although with the present data set
we cannot clearly identify a statistically significant feature.
Finally, we used the Beloborodov~(2013) model to fit the hard-band spectrum and show
that the model successfully describes the observed spectrum, and can be used to constrain
the emission geometry; the angle between the rotation and magnetic axes of the neutron
star is inferred to be $\sim$20$^\circ$, and the angle between the rotation axis and line-of-sight
is inferred to be $\sim$50$^\circ$ (\cite{ahk+13}).

\subsection{1E~2259$+$586}
1E~2259$+$586 is one of the brightest known magnetars\footnote{See the online magnetar catalog for a compilation
of known magnetar properties, http://www.physics.mcgill.ca/$\sim$pulsar/magnetar/main.html} (\cite{ok13})
in the soft band ($\lapp$10 keV) despite the fact that its magnetic-field strength is relatively
low ($B=5.9\times 10^{13}\ \rm G$). However, in the hard band
it was detected only marginally with {\em RXTE} (\cite{khh+06}). With the marginal detection,
Kuiper et~al. (2006)
measured the hard-band photon index for the pulsed emission to be $-1.02 \pm 0.19$,
extremely hard. If so, we should be able to clearly detect the source
with {\em NuSTAR} to first verify the previous results for the pulsed emission and to measure the
total spectrum in the hard band for the first time. With a clear detection, the source will be useful
for testing the hard X-ray emission model of Beloborodov~(2013).

The source was observed with {\em NuSTAR} in 2013 Apr--May for an exposure of 110 ks.
We detected the source above the background level up to $\sim$25 keV, and used an absorbed blackbody
plus broken power-law model to fit the spectrum. Interestingly, the hard-band spectrum for the total
emission ($\Gamma \sim 1$) is not as hard as the pulsed one. Further analysis is on-going to measure other
source properties including the pulsed spectrum (\cite{v14}).

\subsection{Geminga}
Geminga is one of the brightest RPPs in the gamma-ray band. In the soft X-ray band, it is faint and shows
a blackbody plus power-law spectrum. This source can serve as an archetype for many
other X-ray faint/gamma-ray bright RPPs (see Kaspi, Roberts, \& Harding 2006, for a review).
The power-law spectral component is not very well constrained,
and does not seem to connect to the gamma-ray band, suggesting a possible spectral break between the
hard X-ray and gamma-ray band (\cite{kpz+05}). With {\em NuSTAR}, we can measure the hard X-ray
spectrum very well, and constrain the non-thermal spectral component. Furthermore, we will be able to
detect the spectral break if it is in the {\em NuSTAR} band.

The source was observed in 2012 Sep with {\em NuSTAR} for $\sim$50 ks. Archival {\em XMM-Newton} and
{\em Chandra} data were combined with the {\em NuSTAR} data for the spectral analysis. From the spectral
analysis, we find that the non-thermal X-ray spectrum seems to extend up to the {\em Fermi} band, implying no
spectral break in the hard X-ray band. However, combining the soft-band data together was not trivial
for an unabsorbed source like this; cross-calibration issues in the soft-band seem to be severe.
As the soft-band spectra affect the hard-band results in the analysis, we are currently conducting
careful sensitivity studies (\cite{m+14}).

\subsection{AE Aquarii}
AE Aquarii is a very interesting IP which was suggested to accelerate particles in its magnetosphere as
RPPs do based on a {\em Suzaku} detection of a non-thermal spectral component and a spiky pulsation
in the hard band (\cite{thi+08}). However, the hard-band detection was marginal and required confirmation.

The source was observed by {\em NuSTAR} in 2012 Sep, and
was clearly detected up to $\gapp$20 keV. We find that the spectrum of the source can be described
with a multi-temperature thermal model or a thermal plus non-thermal model; we cannot clearly rule out
the thermal model in the hard X-ray band ($\gapp$10 keV).
Furthermore, we find no evidence of the spiky structure in the hard-band
pulse profile seen in the {\em Suzaku} observation (\cite{thi+08}). Therefore, we conclude that AE Aquarii
is more likely an accretion-powered IP than a pulsar-like accelerator (\cite{kab+14}).
However, we find that the highest
temperature of the thermal model is significantly lower than those of other IPs; conventional accretion
shock emission models seem not to work for this source. Therefore, a new model will be applied
to interpret the spectrum we measured (\cite{kab+14}).

\subsection{PSR~J1023$+$0038 and 1E~1048$-$5937}
PSR~J1023$+$0038 is a millisecond radio pulsar in an LMXB which
has shown a hard power-law tail ($\Gamma \sim 1.3$) in the previous {\em XMM-Newton}
and {\em Chandra} observations (\cite{akb+10}; \cite{bah+11}).
It can serve as a template for understanding non-thermal emission from quiescent LMXBs which are
suggested to have unseen millisecond radio pulsars.

If the non-thermal spectrum measured in the soft band extends to the {\em NuSTAR} band, we will accurately
measure the spectrum with {\em NuSTAR} and compare it with those of other similar quiescent LMXB systems.
We observed the source with {\em NuSTAR} in 2013 June for 100 ks, and detected the orbital modulation up
to $\sim$50 keV with {\em NuSTAR}, and are presently conducting
detailed spectral/temporal analyses (\cite{t14}).

1E~1048$-$5937 is a magnetar with spin-inferred magnetic-field strength of $\sim$4$\times10^{14}\ \rm G$.
However, hard-band emission from the magnetar 1E~1048$-$5937 has not yet been clearly detected. Our goal for observing
this source is to first detect the hard-band emission, and test the putative
correlation between the degree of spectral turnover ($\Gamma_{\rm s} - \Gamma_{\rm h}$) and the spin-inferred $B$
reported by Kaspi \& Boydstun (2010).
The correlation suggests that {\em NuSTAR} should be able to clearly detect the source in the hard band
with a $\sim$400-ks exposure.

We detected the hard-band emission up to $\gapp$20 keV with {\em NuSTAR}. More interestingly, there were several
bursts detected during the observations. We are currently analyzing the data to study the bursts and the
persistent properties of the source (\cite{a14}; \cite{ak+14}).

\section{Conclusions}
\label{sec:concl}
Since its launch in 2012 June, {\em NuSTAR} has performed excellently. As a part of {\em NuSTAR}'s science program,
we have observed carefully selected magnetars, RPPs and a white dwarf. Many interesting results are ready to be
published for some sources, and detailed data analyses are on-going for others.

\acknowledgements
This work was supported under NASA Contract No. NNG08FD60C, and  made use of data from the {\it NuSTAR} mission,
a project led by  the California Institute of Technology, managed by the Jet Propulsion  Laboratory,
and funded by the National Aeronautics and Space  Administration. We thank the {\it NuSTAR} Operations,
Software and  Calibration teams for support with the execution and analysis of  these observations.
This research has made use of the {\it NuSTAR}  Data Analysis Software (NuSTARDAS) jointly developed by
the ASI  Science Data Center (ASDC, Italy) and the California Institute of  Technology (USA). V.M.K. acknowledges support
from an NSERC Discovery Grant, the FQRNT Centre de Recherche Astrophysique du Qu\'ebec,
an R. Howard Webster Foundation Fellowship from the Canadian Institute for Advanced
Research (CIFAR), the Canada Research Chairs Program and the Lorne Trottier Chair
in Astrophysics and Cosmology. A.M.B. acknowledges the support by NASA grants NNX10AI72G and NNX13AI34G.
Part of this work was performed under the auspices of the U.S. Department
of Energy by Lawrence Livermore National Laboratory under Contract
DE-AC52-07NA27344.

\end{document}